# Specification-Based Code–Text–Code Reengineering for LLM-Mediated Software Evolution

Oleg Grynets
EPAM Systems
McLean, Virginia, USA
oleg_grynets@epam.com

Vasyl Lyashkevych
EPAM Systems
Lviv, Ukraine
vasyl_lyashkevych@epam.com

Arsen Dolichnyi
EPAM Systems
Lviv, Ukraine
arsen_dolichnyi@epam.com

Roman Piznak
EPAM Systems
Lviv, Ukraine
roman_piznak@epam.com

Taras Zelenyy
EPAM Systems
Lviv, Ukraine
taras_zelenyy@epam.com

Volodymyr Morozov
EPAM Systems
Kyiv, Ukraine
volodymyr_morozov@epam.com

***Abstract*—Large language models (LLMs) increasingly support software migration, refactoring, documentation, program explanation, and source code regeneration. However, direct code-to-code transformation remains challenging to control because it can preserve surface-level syntax while introducing semantic drift, hidden behavioral changes, loss of traceability, non-idiomatic target implementations, or incomplete reconstruction of domain logic. This paper proposes a specification-based Code–Text–Code reengineering framework for LLM-mediated software evolution. The central idea is to transform source code into a neutral textual specification that captures program behavior, identifiers, computational flow, conditions, side effects, data dependencies, and domain-specific intent without directly transferring the source language syntax. This specification is then used as a controlled intermediate representation for regeneration, migration, or modification of the target code. The proposed framework combines factual context extraction, Code2Text generation, iterative verification between source code and text specification, optional "human in the loop" correction, Text2Code generation, target code verification, retrieval-augmented grounding and semantic-aware chunking, and transformation loss estimation. The knowledge representation layer integrates metadata derived from AST, graph-based dependency structures, neutral natural language specifications, technical documentation, business documentation, and architecture-level representations. The conducted experiments include a Code–Text–Code dataset built from multiple programming languages and SQL dialects, comparison of intermediate representations, retrieval evaluation, documentation transformation evaluation, and prompt tuning using DSPy/MIPROv2. A graph formalization using structural preservation, reverse compatibility, interface stability, and total graph similarity is implemented to estimate transformation losses. The results support the interpretation of the Code–Text–Code approach not as a simple code transformation, but as a controlled specification-based reengineering process for LLM-mediated software evolution.**

## I. Introduction

LLMs have fundamentally changed the way software systems are created, modified, documented, and migrated [1]–[3]. Modern software engineering practices increasingly use LLMs for code generation, code completion, code explanation, migration between programming languages or database dialects, test generation, defect localization, technical documentation, and repository-level assistance [1]–[5]. Recent surveys confirm that code generation and software engineering are among the most active areas of LLM application, while highlighting unresolved issues related to evaluation, reliability, context management, and the quality of generated code.

A common approach to LLM modernization is direct code-to-code conversion. In this mode, source code written in one programming language, framework, or database dialect is provided to a model, which then has to generate equivalent target code. While this approach is convenient, it has several limitations.

First, direct conversion may preserve syntactic patterns from the source language, even if they are not idiomatic or inappropriate in the target environment.

Second, the model may implicitly infer missing information and implement additional behavior that was not present in the original system.

Third, the generated code may appear syntactically correct, while partially losing business semantics, data dependencies, exception handling logic, or non-functional assumptions.

Fourth, direct transformation offers limited traceability between the source code, model interpretation, and the regenerated artifact.

These limitations suggest that LLM-mediated software evolution should not be viewed as code generation alone. Instead, it should be interpreted as a reengineering process in which software artifacts are moved through multiple levels of representation.

This paper focuses on specification-based Code–Text–Code reengineering as one such representation-driven software evolution process. The key idea is to avoid uncontrolled direct code-to-code transformations and to introduce an intermediate, neutral textual specification between the source and target code. This specification acts as a controlled representation of the behavior of the program, separating what the software does from how it is implemented in a particular language, framework, or platform. This creates a need for representation layers that connect code, documentation, architecture, metadata, and regenerated artifacts in a repeatable transformation cycle [6]. In migration scenarios, the same need is reinforced by fine-tuned LLM-based code migration frameworks that combine feature detection, transformation rules, expert feedback, and iterative evaluation [7]. This need is also consistent with the broader interpretation of intelligent monitoring as an information technology for context-aware decision-making strategy selection [8].

In this view, monitoring is not limited to the passive collection of technical indicators, but supports the selection of appropriate decisions and response strategies under changing contextual conditions [8]. Therefore, specification-based Code–Text–Code reengineering can be considered not only as a transformation pipeline, but also as an object of intelligent monitoring, where semantic consistency, structural preservation, interface stability, and transformation losses must be continuously assessed during LLM-mediated software evolution.

The proposed approach follows a transformation scheme:

$$C_0 \rightarrow S_n \rightarrow C_1, \quad (1)$$

where $C_0$ is the source code, $S_n$ is a neutral text specification, and $C_1$ is the regenerated, migrated, or modified target code. Unlike informal comments or ordinary documentation, $S_n$ is considered a structured and verifiable artifact of knowledge representation. It should preserve behavioral intent, identifiers, control flow, conditions, side effects, input/output assumptions, and domain-level semantics. At the same time, excessive technical jargon and language-specific constructs that can distort the generation of target code should be avoided.

This interpretation is particularly important for systems modified or created using LLMs. Such systems may evolve through iterative cycles of explanation, documentation, requirements updates, code generation, testing, and deployment. Thus, software evolution becomes not only a sequence of code changes but also a sequence of representation transformations. This creates a need for mechanisms that can control semantic consistency, structural preservation, interface stability, traceability, and transformation losses at different representation levels.

Fig. 1 summarizes this idea by showing how source code elements are progressively transformed into technical documentation, high-level functional descriptions, business documentation, and requirements. The figure emphasizes that Code–Text–Code reengineering is not a single direct translation step, but a multi-level transformation across code, technical, and business representations.

The figure also highlights the role of requirements and documentation as intermediate artifacts that can later guide reverse transformation from updated business intent back to technical documentation and regenerated code.

This research was motivated by practical experiments with code-text-code pipelines, intermediate representations, retrieval-augmented generation, semantic-aware chunking, documentation transformation, project graph representation, and graph-based similarity scoring. The experimental evidence shows that a neutral natural language representation can serve as a useful intermediate layer, while AST-derived metadata, graph representations, retrieval mechanisms, and validation loops are necessary to maintain traceability and reduce semantic drift. In the developed prototype, the Code–Text–Code architecture implements context extraction, Code2Text transformation, iterative validation, optional human intervention, Text2Code generation, and target-code validation as parts of a controlled pipeline.

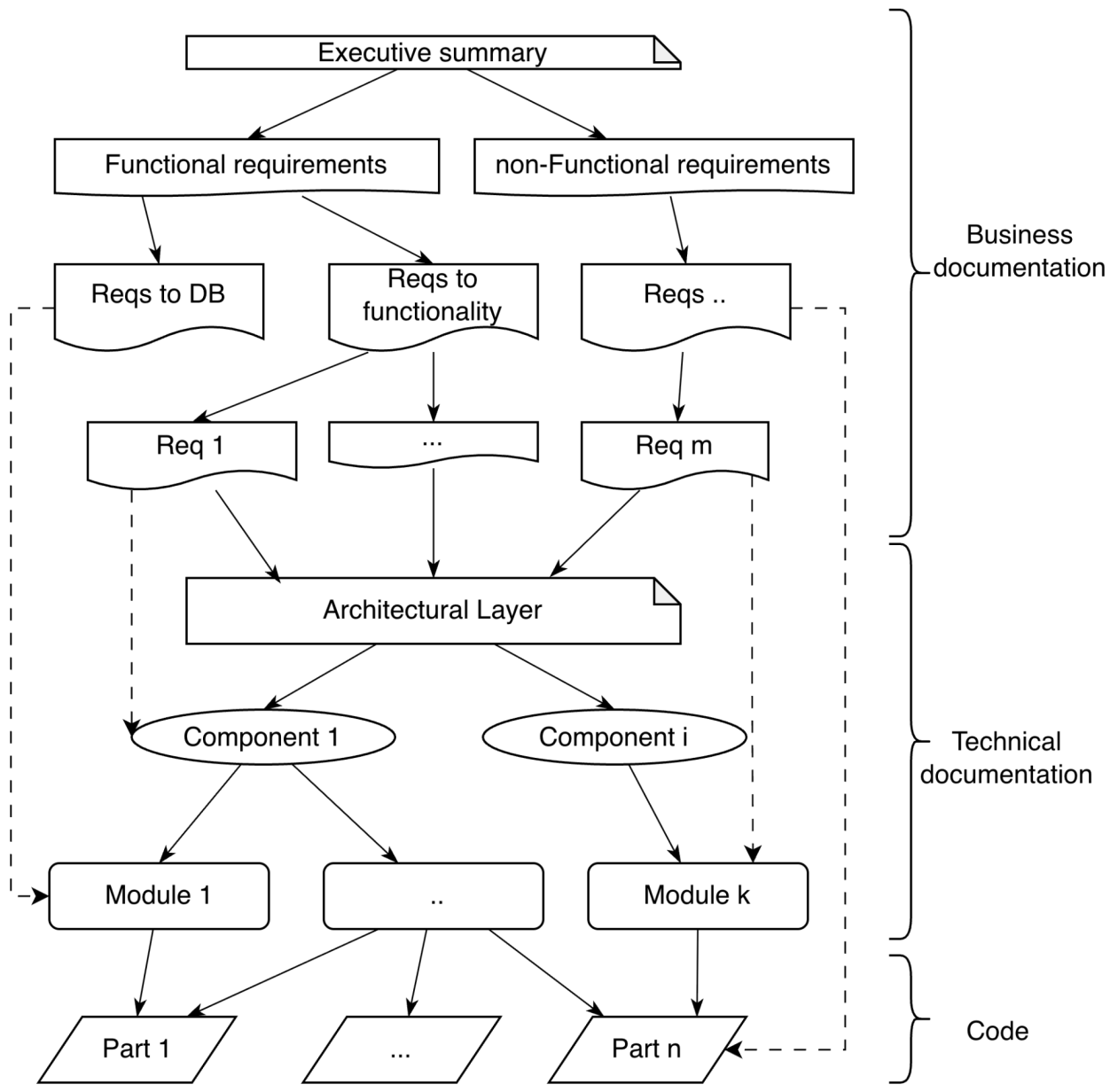

Fig. 1. Specification-based Code–Text–Code reengineering model showing the transition from code elements to functions, components, high-level technical documentation, business documentation, and requirements.

The purpose of this paper is to define a specification-based "Code-Text-Code" reengineering framework for LLM-mediated software evolution and demonstrate how such a framework can integrate textual specifications, metadata, graphs, retrieval mechanisms, validation loops, and formal metrics to assess transformation losses.

The main achievements of this paper are as follows:

- A conceptual interpretation of the "Code-Text-Code" transformation as a specification-based software reengineering process rather than a simple code translation pipeline.
- A pipeline architecture that includes factual context extraction, neutral textual specification generation, iterative validation, human-in-the-loop correction, target-code generation, target-code validation, and DBMS-supported feedback.
- A knowledge representation model that combines AST-derived metadata, ontology storage, graph-based dependencies, neutral textual specifications, technical documentation, business documentation, and architecture-level views.
- An experimental evaluation perspective based on intermediate representation comparison, retrieval quality, documentation transformation quality, prompt tuning, semantic consistency, and transformation-loss estimation.
- A graph-based formalization for estimating transformation losses between source and target software structures using structural preservation, reverse compatibility, interface stability, and total graph similarity.

Taken together, these contributions define Code–Text–Code reengineering as a representation-driven process in which neutral textual specification is the central control artifact, while metadata, ontology, graph structures, retrieval, validation, and DBMS feedback provide supporting evidence and quality control.

## II. Related Work

### A. LLMs in Software Engineering

LLMs have become an important technology for software engineering tasks, including code generation, code summarization, defect detection, program debugging, code review, test generation, and requirements support. Surveys by Hou et al. [1], Zhang et al. [2], and Jiang et al. [3] show that LLM-based software engineering research has moved from function-level code generation toward repository-level reasoning, multi-file editing, and software engineering automation. These surveys also reveal recurring problems: insufficient evaluation, hallucinated functionality, context length limitations, security risks, and the difficulty of aligning generated code with real-world project constraints.

Code translation and migration are closely related to this work. Traditional code translation attempts to map constructs from the source language to the target language. LLM-based translation adds flexibility, but also introduces new risks: the translated code may be syntactically correct while preserving source language idioms, changing behavior, or ignoring platform-specific assumptions. This is especially important for database migration, where semantic equivalence depends on data types, transaction semantics, stored procedures, SQL dialect-specific constructs, and exception handling [7], while broader architecture-level representation is required to preserve dependencies across code, documentation, and regenerated artifacts [6].

The relevance of LLM for software evolution is illustrated by benchmarks such as SWE-bench, which assesses whether models can solve real-world problems on GitHub. SWE-bench contains 2294 software engineering problems from 12 Python repositories and shows that solving real-world problems requires understanding and coordinating changes across functions, classes, and files [4]. This supports the argument that software changes mediated by LLM cannot be reduced to generating isolated code snippets; they require project-level representation, context discovery, dependency awareness, and validation.

### B. LLM-Based Code Generation and Code Translation

Code generation using LLMs has been investigated using benchmarks such as HumanEval, Mostly Basic Programming Problems (MBPP), BigCodeBench, ClassEval, and repository-level tasks [9]. Recent studies highlight that LLMs can generate syntactically plausible code, but still fail to cope with hidden semantic requirements, complex dependencies, security constraints, or real-world integration assumptions [3], [5].

Open code-oriented foundation models, such as Code Llama, also demonstrate that code generation has become a specialized direction of LLM development rather than only a general text generation task [10].

Code infilling models such as InCoder further show that code generation often requires reasoning over incomplete program contexts, missing spans, and surrounding code structure [11].

Large open code models such as StarCoder extend this trend by training on large-scale permissively licensed code corpora and supporting multilingual code generation scenarios [12].

At the same time, studies of how programmers interact with code-generating models show that developers often use such tools iteratively, checking, correcting, and grounding generated suggestions in their own understanding of the task [13].

Earlier work on program synthesis with large language models also shows that LLM-based code generation can be interpreted as a form of probabilistic program synthesis, where generated solutions require systematic evaluation rather than blind acceptance [14].

These works show that LLM-based code generation has progressed from isolated benchmark tasks to specialized code models, infilling, program synthesis, and developer-in-the-loop workflows. However, they also reinforce the need for explicit intermediate representations, validation mechanisms, and traceability when generated code is used for software evolution rather than isolated programming tasks.

The work "NL in the Middle" directly motivates this paper, investigating whether LLM-based code translation can benefit from intermediate representations, including natural language annotations and ASTs [15]. The authors report that intermediate representations can guide the translation process, but also show that the quality of the intermediate representation strongly influences the final translation results. This is consistent with the central assumption of our work: a neutral textual specification can act as a controlled semantic mediator between the source and target code.

### C. Code Summarization and Code-to-Text Transformation

Traditionally, the goal of code summarization has been to create short, natural language descriptions of functions, methods, classes, or files [16]. With LLM, code summarization has expanded beyond simple comments to include technical documentation, business explanations, API descriptions, and project-level summaries. Recent work on code summarization beyond the function level highlights the need to summarize larger code artifacts and assess the quality of summaries beyond surface-level metrics [17].

However, simple code summarization is not sufficient for code-text-code reengineering. A summary may miss details that are important for regeneration. In contrast, a neutral textual specification should capture behavior, conditions, inputs, outputs, side effects, dependencies, and constraints. It is closer to a controlled semantic specification than to an explanatory comment. Therefore, the proposed framework considers Code2Text as a specification recovery task, not a generalization task.

### D. Intermediate Representations for Code Transformation

Intermediate representations (IRs) are widely used in compilers, static analysis, program transformation, and software understanding. Traditional IRs include ASTs, control flow graphs, data flow graphs, program dependency graphs, bytecode representations, compiler-style IRs, and domain-specific representations. In LLM-mediated code transformation, IRs can provide structured guidance and reduce ambiguity, continuing earlier work on structured code representations such as code2seq [18]. Similarly, code2vec demonstrated that distributed representations of code can be learned from structural paths rather than from raw tokens alone [19].

Recent work on intermediate representations for LLM-based code translation compares direct translation with representations such as natural language, pseudocode, and AST-based forms [15]. In the present experimental setting, this comparison was extended to pseudocode, natural-language IR, graph-based IR, and compiler-style IR. Each has different strengths. ASTs preserve syntactic structure but are language-specific. Compiler IRs provide an accurate low-level representation but can be too far removed from business semantics. Graph IRs are useful for dependencies but require robust parsing. Natural-language IR is less formal but better aligned with LLM reasoning and human validation.

The external study in [15] motivates the use of natural-language intermediate representations for code translation. The Code-to-Natural-Language pipeline, in our experiments, was also the strongest among the tested variants, which supports using neutral text as the central intermediate specification layer.

### *E. Program Specifications and Specification Generation*

Specification-based software engineering uses requirements, contracts, preconditions, postconditions, invariants, formal annotations, and behavioral descriptions to support validation and verification. Recent studies of LLM-based specification generation show that LLMs can generate useful formal specifications from code, but correctness and verifiability remain challenging. For example, SpecGen uses LLMs to generate formal program specifications and reports verifiable specifications for 279 out of 385 programs in its evaluation [20].

This direction is important for this work because it shows that LLMs can support the transition from code to specification. However, our focus is broader: the generated specification is not necessarily formal in the mathematical sense; it is a neutral textual specification designed to support supervised regeneration, human validation, RAG grounding, graph tracking, and transformation loss estimation.

### *F. Graph-Based Program Representation*

Graph representations are central to static analysis and software understanding. The Code Property Graph (CPG), proposed by Yamaguchi et al., combines abstract syntax trees, control flow graphs, and program dependency graphs into a single representation [21]. In the proposed framework, graph representation is used not only for vulnerability analysis or static analysis, but also for tracking and estimating transformation losses.

The proposed framework also uses graph IR as a traceability layer and notes that CPG explicitly unifies AST, CFG, and DFG nodes in a single supergraph. This view is also consistent with data-flow-aware code representation models such as GraphCodeBERT, where structural and data-flow information improves program representation beyond plain token sequences [22].

### *G. Ontology representation*

Ontology-based representation can be used as an additional intermediate layer in Code–Text–Code reengineering. While natural-language specification captures the behavioral meaning of the source code, ontology represents this knowledge in a structured and reusable form: domain entities, code objects, functions, data structures, dependencies, constraints, inputs, outputs, and transformation rules. This makes the intermediate representation not only readable for humans and LLMs, but also suitable for retrieval, validation, reasoning, and traceability [23]–[25].

In this case, ontology helps connect source code, metadata, graph structures, textual specifications, technical documentation, and regenerated code into one coherent knowledge model. Earlier work on intelligent diagnostics showed that software ontology can represent a subject domain through concepts and relations to support reasoning and diagnostic task solving [23]. The same principle can be applied to specification-based Code–Text–Code reengineering: raw technical artifacts are transformed into structured knowledge representations that can be searched, interpreted, validated, reused, and monitored during LLM-mediated software evolution.

### *H. Retrieval-Augmented Generation for Code*

The retrieval component of the proposed framework also continues an earlier line of research on specialized search services for intelligent diagnostic tasks. In previous work, a search service was proposed for retrieving diagnostic information required to solve tasks related to computer means [26]. In the present study, this idea is extended from diagnostic information retrieval to LLM-mediated software reengineering. Retrieval is used not only to find relevant information, but also to ground Code2Text and Text2Code transformations with evidence from source code, metadata, documentation, dependency graphs, and external technical knowledge [26]–[30]. This role is also supported by pre-trained programming-language and natural-language models such as CodeBERT, which connect code and natural language representations in a shared semantic space [31].

Retrieval-augmented code generation has become important for repository-level tasks, as LLMs often require project context, library documentation, API definitions, examples, and related code snippets. Recent studies on code generation with retrieval-augmented generation classify the field by retrieval method, generation strategy, model architecture, training paradigm, and evaluation protocol [27]. CodeRAG and CodeRAG-Bench also show that retrieval can support code completion and code generation at the repository level, but retrieval quality, query construction, and relevance matching remain challenging [28], [29].

In our framework, RAG is used to ground Code2Text and Text2Code transformations. The conducted experiments indicate that retrieval over raw SQL text can lead to noisy results, while structured metadata derived from AST provides semantic understanding of the code structure. Therefore, we advocate a hybrid retrieval that combines vector search, graph traversal, metadata filtering, re-ranking, and verification.

### *I. Semantic-Aware Chunking*

Chunking is a critical preprocessing step for large codebases. Poor chunking can break syntax, remove dependencies, fragment semantic units, or create incomplete context for LLM. In the developed experimental pipeline, improper code chunking was observed to increase the risk of incomplete code coverage, hallucinated behavior, invalid transformations, additional manual correction, and higher processing cost.

Unlike regular document chunking, code chunking must take into account syntax, semantic boundaries, dependency relationships, and recompilation behavior. Therefore, semantic-aware chunking is considered in this article as part of the reengineering methodology, rather than as a minor implementation detail.

### J. Evaluation of LLM-Generated Code

Evaluating code generated by LLM remains challenging. Common metrics include pass@k, unit test success, syntactic validity, exact match, BLEU, CodeBLEU, execution accuracy, human evaluation, LLM as a judge, vulnerability checks, and repository-level performance benchmarks. However, these metrics do not fully account for semantic preservation, traceability, or loss of transformation at different representation levels.

Security research also shows that generated code can contain vulnerabilities, even if it appears to be correct. Recent empirical studies have investigated whether LLMs consider security and how they behave when asked to generate or fix secure code [32], [33]. This supports the need for validation cycles and explicit validation of generated artifacts.

The proposed framework is also connected to classical software reengineering and reverse engineering. Chikofsky and Cross defined reverse engineering and design recovery as processes for analyzing existing software systems and identifying their components and interrelationships at higher levels of abstraction [34]. Specification-based Code–Text–Code reengineering follows the same general idea, but extends it with LLM-mediated representation transformation, neutral textual specification, and regeneration of target code.

Formal specification research also emphasizes the importance of precise behavioral descriptions, requirements, assumptions, and validation criteria for software development and evolution [35]. In the proposed approach, the neutral textual specification is not a full formal specification, but it plays a similar mediating role by making behavior explicit before regeneration.

These classical foundations clarify why the proposed framework should be treated as reengineering rather than only as code generation. The novelty lies in combining design recovery, specification recovery, LLM-mediated transformation, and validation across code, text, documentation, and graph representations.

### K. Research Gap

The reviewed literature demonstrates significant progress in LLM-based code generation, code translation, code summarization, RAG, program graphs, and specification generation. However, three gaps remain.

First, most approaches treat code-to-text, text-to-code, documentation generation, and code migration as separate tasks rather than as a single reengineering cycle.

Second, the intermediate text representation is often treated as a temporary artifact or summary rather than a controlled level of specification.

Third, evaluations often focus on local correctness or comparative performance rather than on the transformation losses in code, text, documentation, graphs, and regenerated code.

These gaps motivate a unified framework in which code-to-text, documentation transformation, retrieval, validation, and text-to-code regeneration are treated as parts of one traceable reengineering cycle rather than as isolated LLM tasks.

This paper addresses these gaps by proposing a specification-based "Code-Text-Code" reengineering framework in which a neutral textual specification functions as a controlled intermediate level of knowledge representation for LLM-mediated software evolution.

## III. Specification-Based Code–Text–Code Reengineering Model

### A. Basic Transformation Model

The proposed framework interprets LLM-mediated reengineering as a controlled transformation between representation layers:

$$C_0 \rightarrow I_C \rightarrow S_n \rightarrow C_1 , \quad (2)$$

where $C_0$ is the source code, $I_C$ is a set of factual context extracted from source code, $S_n$ is the neutral textual specification, and $C_1$ is the generated or migrated target code.

The $C_0$ source code contains implementation-specific syntax, identifiers, control structures, dependencies, external calls, data access patterns, and implicit assumptions. The factual context $I_C$ contains information that should not be lost during the conversion. This includes the names of functions, procedures, classes, tables, columns, parameters, conditions, loops, side effects, I/O objects, and dependencies. The neutral textual specification $S_n$ is the central artifact. It is not simply a natural-language summary. It is a controlled representation of software behavior. It should be language-agnostic, preserve behavior, be traceable to source elements, be human-readable for verification, amenable to LLM-based regeneration, free from unnecessary constructs specific to the source language, and clearly describe conditions, data dependencies, and side effects. The target code $C_1$ is generated from $S_n$, not directly from $C_0$. This separation reduces the probability of copying source-specific idioms into the target language. For example, a direct translation from Python to C++ may emulate Pythonic constructs in C++, while a neutral specification can guide the generation of idiomatic C++ behavior.

The role of each intermediate representation differs depending on whether the transformation requires syntactic precision, semantic abstraction, dependency preservation, or stakeholder-level interpretation. Table I summarizes the main representation types used in the proposed framework.

TABLE I. Comparison of Intermediate Representations for Code–Text–Code Reengineering

| IR type | Main role | Advantages | Limitations | Suggestion |
|---|---|---|---|---|
| AST / unified AST | Syntactic structure | Precise parsing; supports metadata extraction | Language-specific; parser variability | Source-code analysis and evidence extraction |
| Pseudocode IR | Algorithmic abstraction | Human-readable; preserves steps | May retain source-specific logic | Algorithm-level transformations |
| Compiler IR | Low-level transformation | Precise; suitable for compiler tasks | Too low-level for business documentation | Optimization and formal code transformation |
| Graph IR | Dependency representation | Captures calls, reads, writes, flow | Requires reliable relation extraction | Traceability and loss estimation |
| Natural-language IR | Semantic mediation | Human-readable; LLM-aligned | Ambiguity risk | Neutral textual specification |
| Technical documentation | Implementation explanation | Useful for developers | May be verbose or incomplete | Project-level understanding |
| Business documentation | Stakeholder abstraction | Supports requirement updates | May omit technical detail | Business-driven evolution |

The comparison shows that no single representation is sufficient for the whole Code–Text–Code cycle. AST and metadata provide reliable evidence, graph IR supports dependency tracking, and natural-language IR provides the most suitable semantic mediation layer for LLM-based regeneration.

### B. Neutral Textual Specification

The neutral textual specification is the key difference between simple code translation and specification-based reengineering. It plays five roles:

- Semantic abstraction – it describes what the program does rather than how the source language implements it.
- Validation artifact – it can be compared with source code to detect missing behavior, unsupported assumptions, or excessive interpretation.
- Communication artifact – human experts can review and correct it more easily than code or graph structures.
- Generation artifact – it provides the model with a controlled language-agnostic description for target-code generation.
- Traceability artifact – it can be linked to source-code elements, metadata records, graph nodes, technical documentation sections, and generated target-code fragments.

These roles show that the neutral textual specification should not be treated as ordinary documentation. It is a controlled intermediate artifact that connects source-code evidence, human understanding, LLM-based regeneration, and downstream validation.

The developed Code–Text–Code prototype supports this role by defining a neutral, non-jargon textual description as the canonical intermediary and by emphasizing that the quality and neutrality of this intermediate text strongly influence the quality of the entire pipeline.

### C. Extended Documentation-Based Evolution Model

The basic Code–Text–Code model can be extended to include technical and business documentation:

$$C_0 \rightarrow TD_0 \rightarrow BD_0 \rightarrow BD_1 \rightarrow TD_1 \rightarrow C_1, \quad (3)$$

where $TD_0$ is technical documentation generated from source code, $BD_0$ is business documentation generated from technical documentation, $BD_1$ is updated business documentation, $TD_1$ is updated technical documentation, and $C_1$ is regenerated target code.

This model supports stakeholder-driven evolution. Business users or analysts can update business documentation without directly editing the code. The framework then determines which technical documents and code chunks need to be changed.

This approach is particularly useful in reporting systems, database-driven applications, enterprise workflows, and migration projects where business logic is embedded in SQL procedures, reports, ETL pipelines, or service components.

## IV. Knowledge Representation Layer

### A. Role of Knowledge Representation

Knowledge representation is central to specification-based reengineering. Without an explicit representation layer, the pipeline becomes a sequence of model calls with limited controllability. To make the intermediate representation reusable and verifiable, the framework combines syntactic extraction, semantic annotation, ontology storage, vector representation, rule-based transformation, and validation. This structure is shown in Fig. 2.

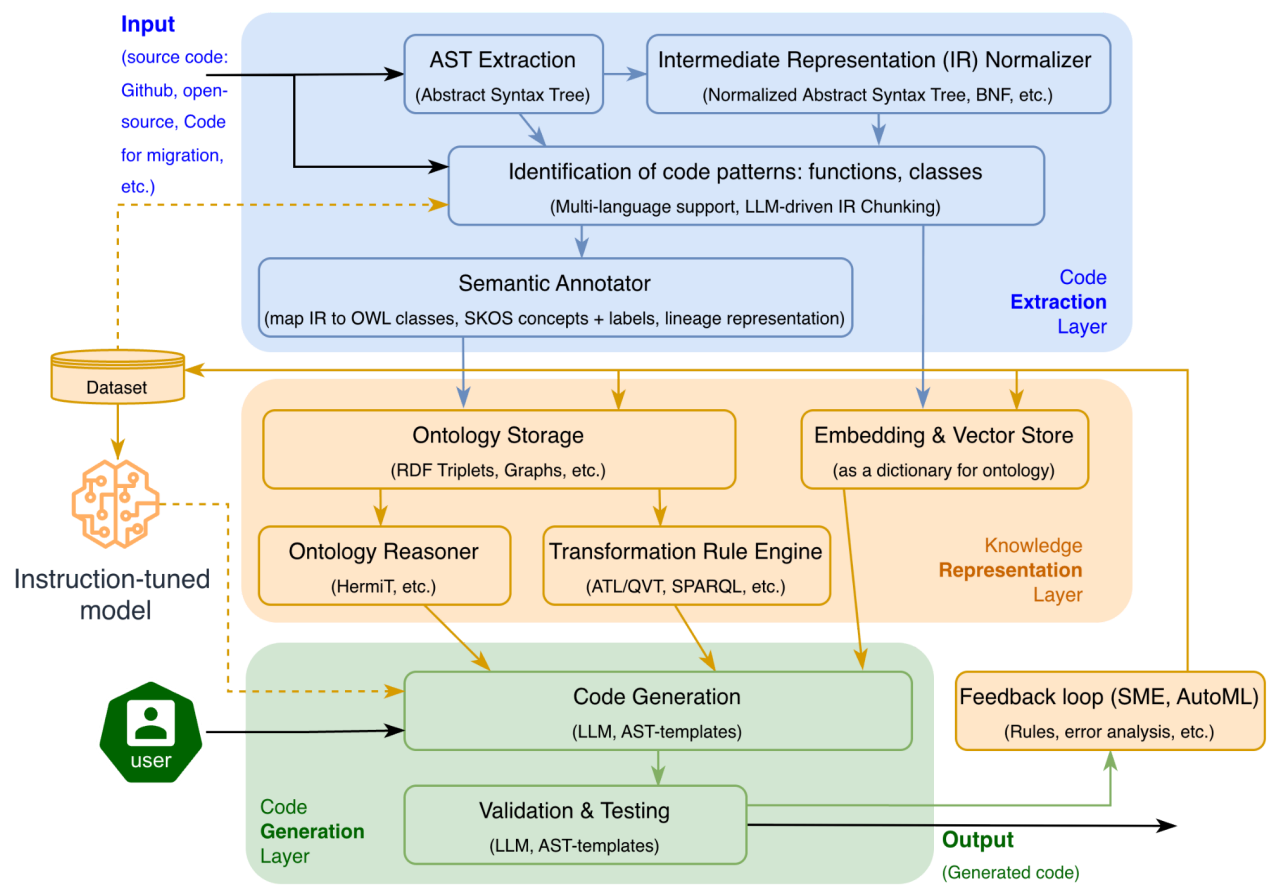

Fig. 2. Ontology-supported Code–Text–Code reengineering architecture with code extraction, IR normalization, semantic annotation, ontology storage, vector storage, reasoning, rule-based transformation, code generation, validation, and feedback loops.

The figure shows that ontology is not used as a replacement for neutral textual specification. Instead, ontology provides a structured semantic layer that supports reasoning, validation, traceability, and reuse of extracted software knowledge.

### B. AST and Metadata Extraction

AST-based analysis is used to extract structured information from code. The metadata extraction pipeline can be represented as follows: "*Code → AST → Metadata → Structured JSON*".

The conducted experiments use this transformation and emphasize that structured metadata extracted from the AST improves semantic understanding of the code structure compared to retrieving raw SQL text.

Metadata can include object type, object name, function or procedure signature, table and column references, input and output parameters, called functions, modified objects, read objects, conditions, error handling, and external dependencies.

### C. Ontology and Semantic Annotation

The ontology-supported architecture maps intermediate representations to OWL classes, SKOS concepts, labels, and lineage relations. Ontology storage can use RDF triples and graph structures, while ontology reasoners and transformation rule engines can support checking, inference, and controlled generation. Fig. 2 shows this architecture as three layers: code extraction, knowledge representation, and code generation. This makes ontology not a replacement for neutral text, but a structured semantic layer that improves retrieval, validation, traceability, and reuse.

## D. Graph Representation

A graph representation captures the relationships between software artifacts. The graph can contain nodes for files, modules, procedures, functions, classes, tables, views, reports, APIs, and external services. Edges can represent read, write, call, import, modify, depend, generate, invoke, or use relationships.

The graph representation supports three important functions: structural context for interpretation, decomposition of large projects, and comparison between source and target systems. By comparing source and target graphs, one can evaluate structural preservation, interface stability, and transformation losses.

## E. Architectural Views

Architectural representations can support specification-based reengineering by organizing knowledge at different levels of abstraction. Architecture frameworks such as TOGAF define views and artifacts for representing data entities, services, components, high-level relationships, standards, protocols, and technologies [36]. In this study, such views are used only as supporting architectural representations, while the neutral textual specification remains the central intermediate artifact.

In the proposed framework, architectural representations are not the primary artifact, but they provide an additional structure for documentation and traceability. The proposed framework treats code, metadata, graphs, documentation, and regenerated artifacts as complementary knowledge representation artifacts. Their roles and validation targets are summarized in Table II.

TABLE II. KNOWLEDGE REPRESENTATION ARTIFACTS AND THEIR ROLE IN THE FRAMEWORK

| Artifact | Source | Representation form | Main role | Validation target |
|---|---|---|---|---|
| Source code | Repository | Files, functions, SQL objects | Executable baseline | Syntax, tests, static analysis |
| AST | Parser | Tree | Structural extraction | Parser success |
| Metadata | AST / analyzer | JSON | Semantic indexing | Completeness and correctness |
| Graph | Metadata + dependencies | Directed graph | Traceability and comparison | Dependency preservation |
| Neutral specification | Code2Text | Controlled text | Semantic mediator | Source-code coverage |
| Technical documentation | Code + graph | Structured text | Developer understanding | Groundedness |
| Business documentation | TechDoc | Stakeholder text | Business-level evolution | Coverage and no hallucination |
| Target code | Text2Code | Code | Regenerated implementation | Spec compliance and tests |

The table shows that validation must be representation-specific. Source code requires syntax and test validation, metadata requires completeness checks, graphs require dependency preservation, and textual specifications require semantic coverage and grounding.

# V. PIPELINE ARCHITECTURE

## A. Overview

The proposed pipeline consists of interconnected transformation and validation stages. At the system level, the framework combines forward transformation from code to documentation and backward transformation from documentation to regenerated code. Fig. 3 presents this end-to-end view, while Fig. 4 complements it with a container-level architecture showing UI, backend orchestration, code processing, metadata extraction, vector database, RAG, LLM gateway, prompt templates, and validation gateway.

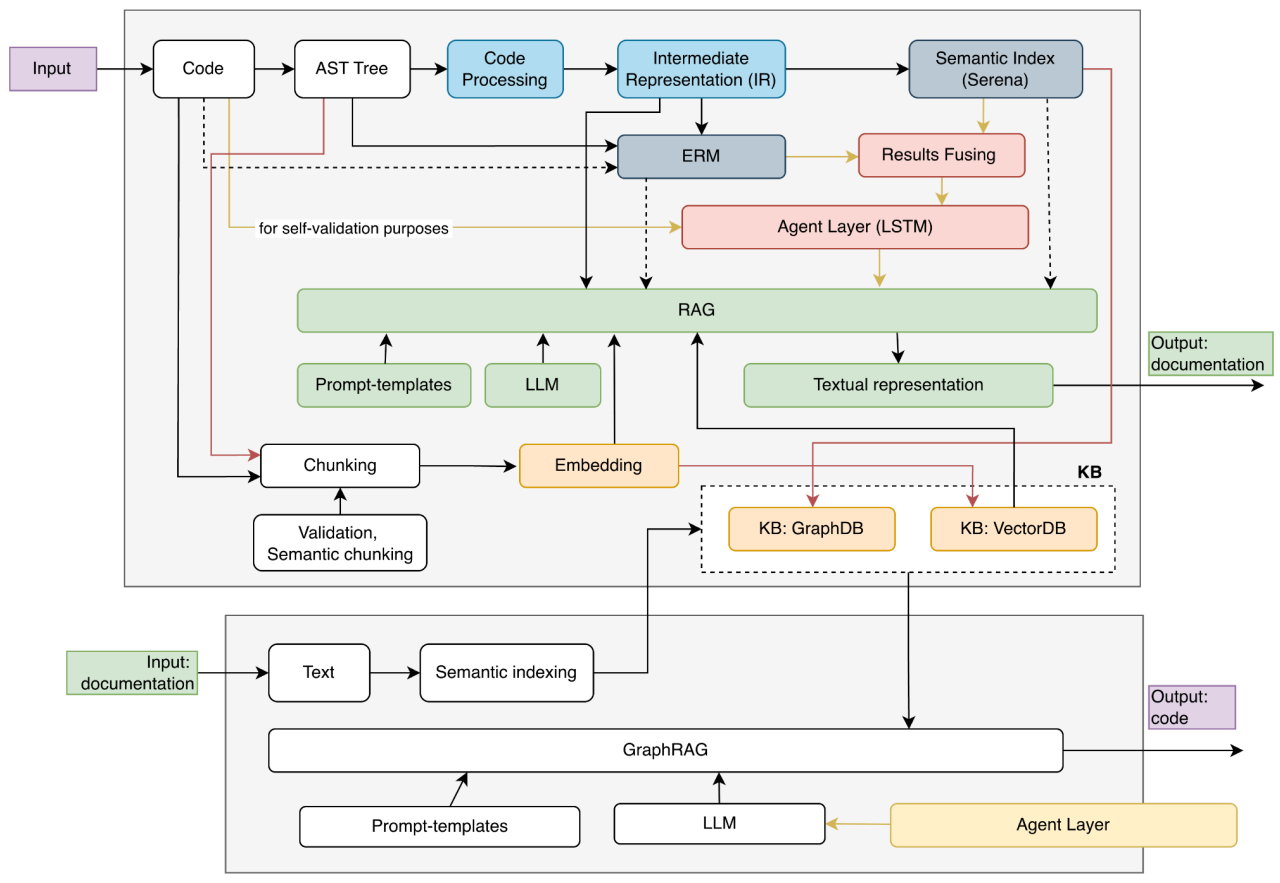

Fig. 3. End-to-end Code–Text–Code architecture connecting code processing, intermediate representation, ERM, semantic index, RAG, GraphDB, VectorDB, textual representation, GraphRAG, and regenerated code.

This architecture separates representation construction from generation. Code processing, IR construction, ERM, semantic indexing, and knowledge bases create the evidence layer, while RAG, GraphRAG, prompt templates, LLMs, and agents support controlled documentation generation and code regeneration. The implementation-level organization of these components is shown in Fig. 4 as a container-level architecture. This view clarifies how the framework can be implemented as a set of interacting modules rather than as a monolithic LLM prompt. It also shows where parsing, embedding, retrieval, prompt loading, LLM calls, and validation are executed.

## B. Factual Context Extraction

Factual context is extracted before the neutral specification is created. Its purpose is to reduce omissions and hallucinations. Context can include identifiers, variable names, function names, database object names, input parameters, output parameters, conditions, loops, side effects, read/write operations, and external dependencies. The key rule is that the context should be factual. It should not introduce interpretation or assumptions. Its role is to constrain the Code2Text stage.

## C. Code2Text Generation

The Code2Text module generates a first version of a neutral text specification. The model receives source code, factual context, metadata, retrieved evidence, and prompt instructions. The output should describe the behavior of the program regardless of the language. Typical Code2Text output should include:

- purpose;
- input data;
- output data;
- main processing steps;
- conditions and branches;
- data access operations;
- side effects;
- error handling;
- dependencies;
- assumptions;
- excluded behavior.

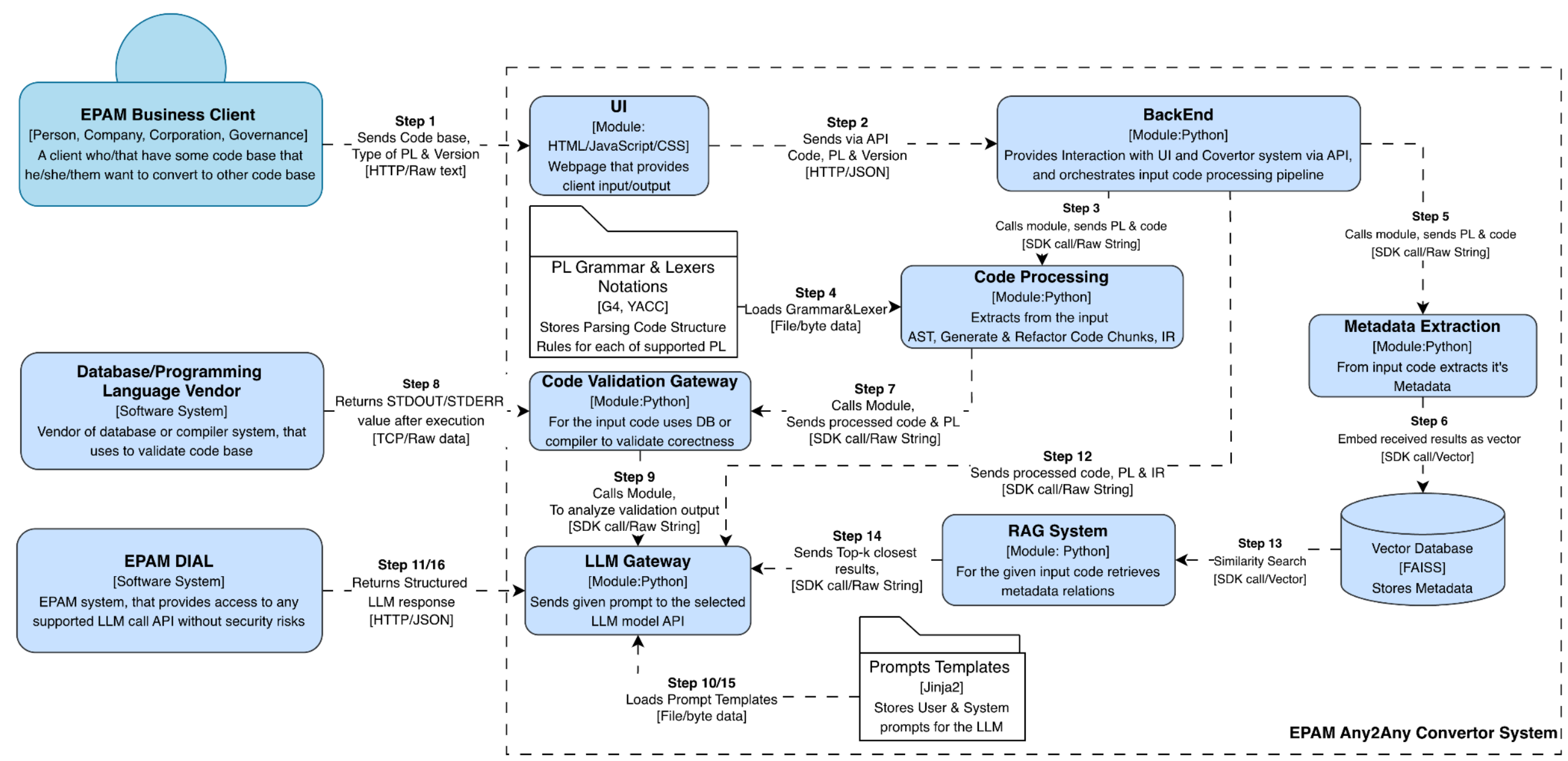

Fig. 4. Container-level architecture of the Code–Text–Code system with user interface, backend orchestration, code processing, metadata extraction, vector database, RAG system, LLM gateway, prompt templates, and validation gateway.

## D. Text Refinement

The refinement module applies minimal targeted changes. This principle is important because full rewriting can lead to new errors. The system should only change those segments that are identified as incomplete or incorrect. This is consistent with the developed Code–Text–Code prototype design, where discrepancies result in targeted refinements rather than complete rewriting.

## E. DBMS Feedback and Execution Validation

For SQL and database migration scenarios, the generated code should be validated not only by syntax checkers but also by DBMS feedback environments. Fig. 5 shows the corrected AnySQL-to-AnySQL experimental pipeline. It combines an extraction layer for dialect and feature detection, static AST/CST parsing for open dialects, SQLGlot or TreeSitter/ANTLR-based parsing when applicable, LLM-based parsing with external documentation retrieval feedback for closed or difficult dialects, and a converter layer that uses AST/CST splitters, DB feedback parsers, static converters, LLM conversion with memory, SQL query refinement, chunk assembly, and final target-dialect validation. For SQL migration, the most critical part of the pipeline is dialect-aware parsing and execution-based validation. This is represented in Fig. 5. The key advantage of this pipeline is that it combines static parsing and LLM-based parsing with DBMS feedback. This reduces the risk of accepting syntactically plausible but non-executable target SQL code.

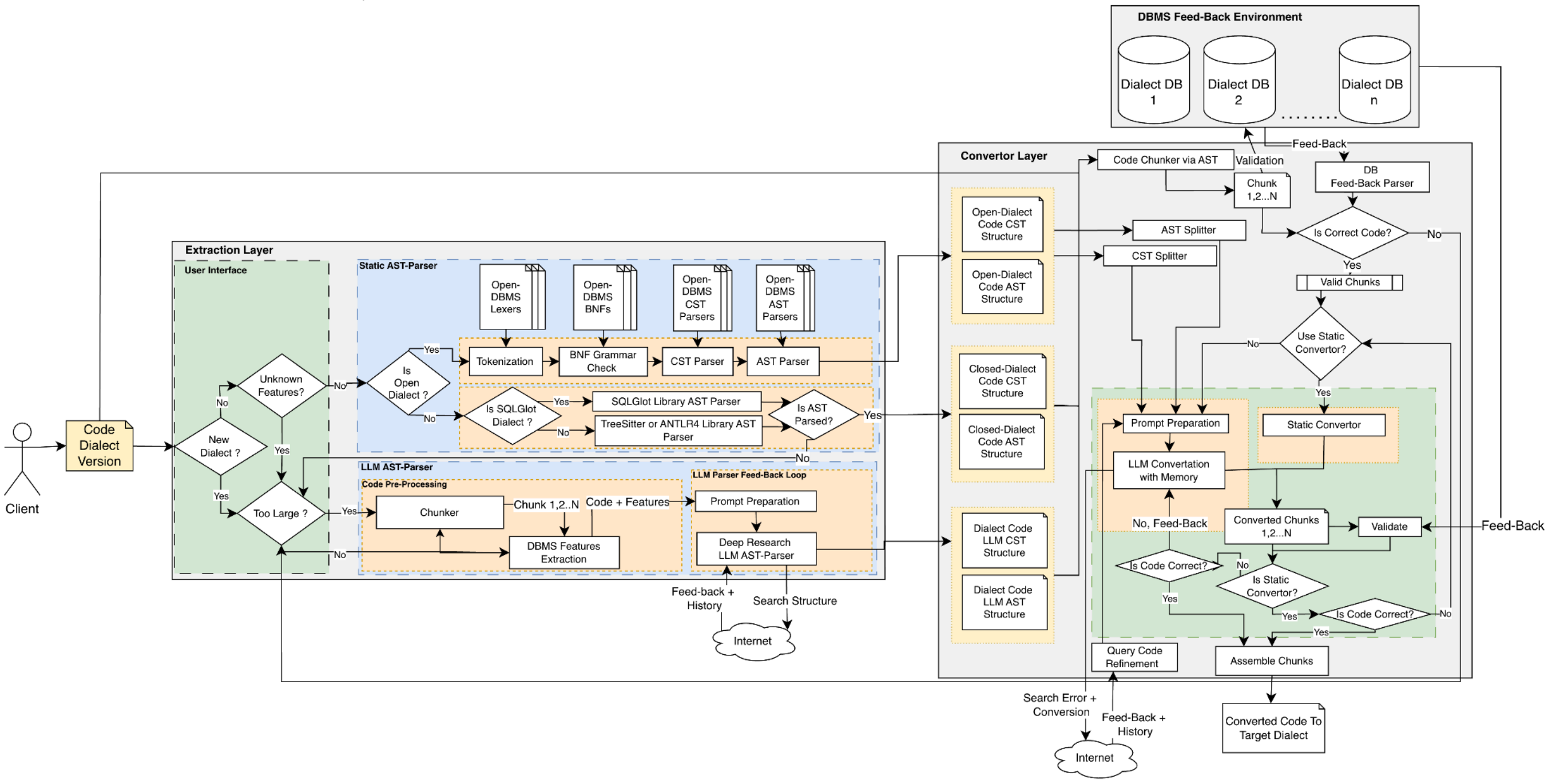

Fig. 5. AnySQL-to-AnySQL experimental conversion pipeline with extraction layer, dialect-aware AST/CST parsing, LLM parser feedback, converter layer, DBMS feedback environment, static or LLM-based conversion, chunk assembly, and validation.

## F. Source-Code-to-Specification Validation

After the initial specification is created, it is compared with the source code. The validation module checks whether relevant source-code elements are represented, whether all conditions and side effects are included, whether inputs and outputs are correct, whether unsupported assumptions appear, and whether the specification contains excessive source-language-specific jargon. Detected differences are passed to a refinement module that applies minimal targeted changes instead of wholesale rewriting.

## G. Human-in-the-Loop Correction

Some ambiguities cannot be resolved automatically. The source code may contain ambiguous names, incomplete comments, hidden business assumptions, or database-specific behavior. Human-in-the-loop correction is not intended to replace automation. It is intended as a precautionary measure for high-risk or ambiguous transformations.

## H. Text2Code Generation and Target Validation

The Text2Code module generates target code from a validated neutral specification. The target language, framework, database dialect, or coding style is provided as part of the generation context. The target code must be syntactically correct, idiomatic, behaviorally consistent with the specification, compatible with the conventions of the target platform, free of unauthorized dependencies, and traceable to the specification statements.

The target verification loop verifies that the generated code fully implements the described functionality, does not introduce additional behavior, preserves required identifiers, follows the conventions of the target language, avoids unauthorized dependencies, preserves input and output data, and implements the required error handling.

# VI. Semantic-Aware Chunking and Retrieval

## A. Chunking Problem

Large codebases cannot always be processed as a single input. They need to be divided into chunks. However, naive chunking can break semantic units and degrade the quality of the transformation. The conducted experiments show that improper chunking can lead to incomplete code coverage, increased hallucination risk, invalid or unsafe generated artifacts, lower transformation quality, additional manual correction, and increased processing cost.

## B. Ideal Code Chunk

An ideal code chunk should satisfy the requirements of atomicity, size balance, minimal overlap, self-parsability, semantic awareness, and self-containment. These properties were used as criteria for defining an ideal code chunk in the conducted experiments. Because chunking quality directly affects retrieval, documentation generation, and code regeneration, chunk properties must be explicitly evaluated. Table III summarizes the required chunk properties and corresponding metrics.

TABLE III. Chunk Properties and Evaluation Metrics

| Property / metric | Meaning | Why it matters |
|---|---|---|
| Atomicity | Chunk represents a meaningful unit | Prevents fragmented logic |
| Size balance | Small enough for model, large enough for context | Avoids context loss |
| Minimal overlap | Avoids unnecessary duplication | Reduces inconsistent regeneration |
| Self-parsability | Chunk can be parsed or checked | Supports deterministic validation |
| Semantic awareness | Reassembly preserves behavior | Prevents semantic drift |
| Self-containment | Minimizes unresolved dependencies | Improves Code2Text quality |
| Chunk Certainty % | AST can be obtained from chunk | Measures syntactic validity |
| Syntax Error Rate % | Execution/parsing error rate | Measures invalid chunks |
| ACTL | Average chunk token length | Measures size suitability |
| Count Rate F1 | Matches expected number of chunks | Detects over/under-chunking |
| Boundary F1 | Matches ground-truth boundaries | Measures segmentation quality |

These metrics allow chunking to be evaluated as a methodological component of reengineering rather than as an implementation detail. In particular, self-parsability, semantic awareness, and boundary quality are critical for reducing semantic drift.

## C. RAG-Supported Code-to-Documentation Extraction

The Code-to-Documentation stage requires grounding in structured code evidence rather than relying only on raw source text. It combines AST extraction, ERM, intermediate representation, validation, semantic chunking, embedding, vector storage, prompt templates, and RAG. Fig. 6 presents this RAG-supported extraction flow.

The figure shows that documentation is generated after code has been parsed, semantically chunked, embedded, and connected to retrieval evidence. This supports more grounded Code2Text generation and reduces the probability of unsupported textual statements.

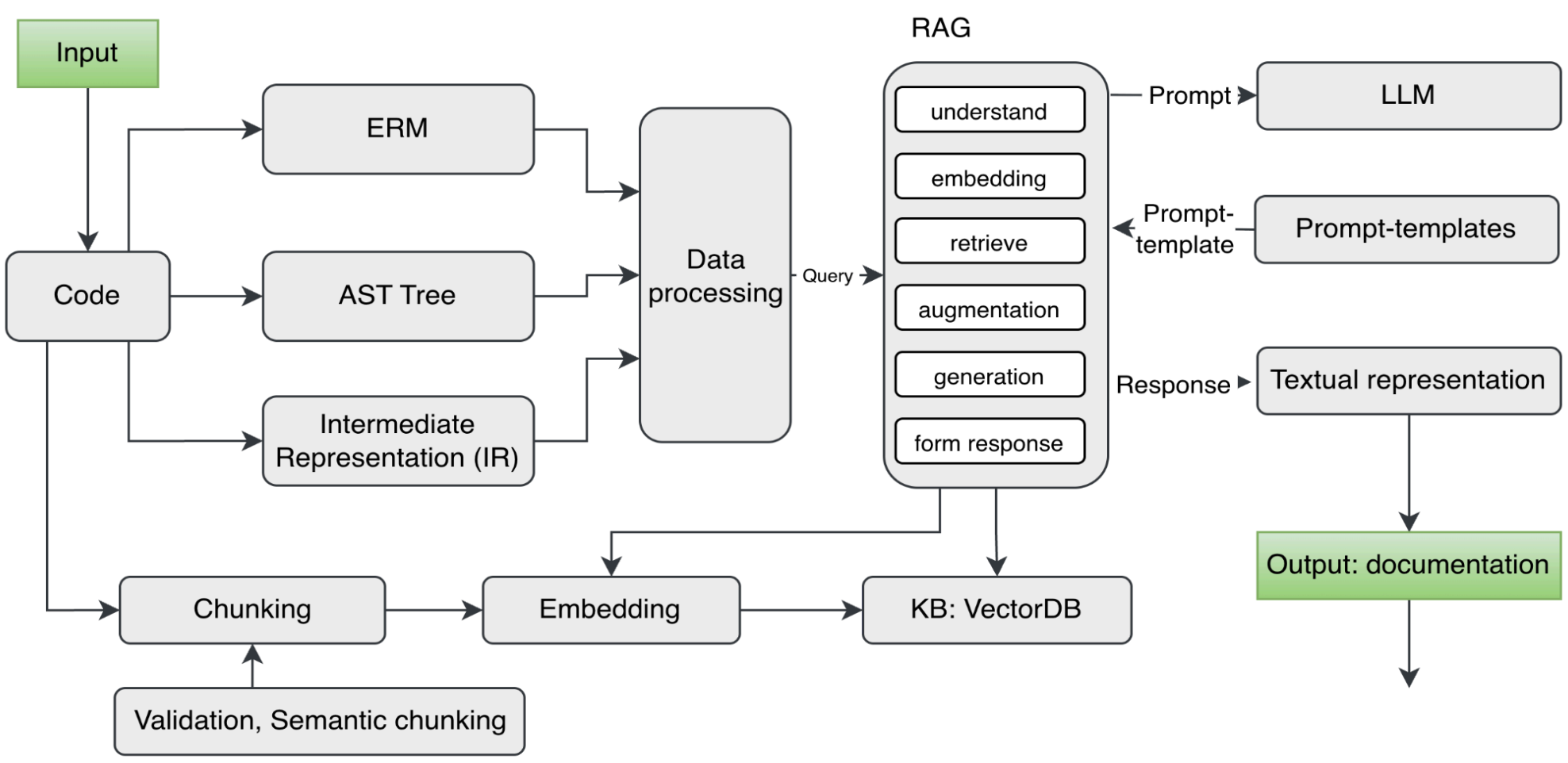

Fig. 6. RAG-supported Code-to-Documentation pipeline with code, AST tree, ERM, intermediate representation, semantic chunking, validation, embedding, vector database, prompt templates, LLM, and textual representation output.

### D. Hybrid Retrieval

Graph retrieval is useful for structural dependencies, relationship traversal, and faster access to related artifacts. The retrieval experiments show that vector retrieval is generally more accurate, while graph search is faster and may be more appropriate when latency is critical: "*Retrieval = VectorSearch + GraphTraversal + MetadataFiltering + Reranking*".

### E. RAG and Internet Search for Code Context

The RAG layer is extended with package descriptions, canonical function or attribute declarations, URLs to documentation or source code, and internet search for third-party libraries and frameworks. Depending on the transformation direction, RAG supports Code2Text by retrieving explanatory context and supports Text2Code by retrieving implementation examples, related chunks, and dependency information.

The reverse direction, from textual requirements or specifications to code chunks, requires mapping textual statements to relevant examples, knowledge-base entries, and cross-language code fragments. This mapping is illustrated in Fig. 7.

This mapping supports controlled Text2Code generation by linking each textual requirement to retrieved evidence and candidate implementation fragments. It also provides a basis for checking whether regenerated code is grounded in the intended textual specification.

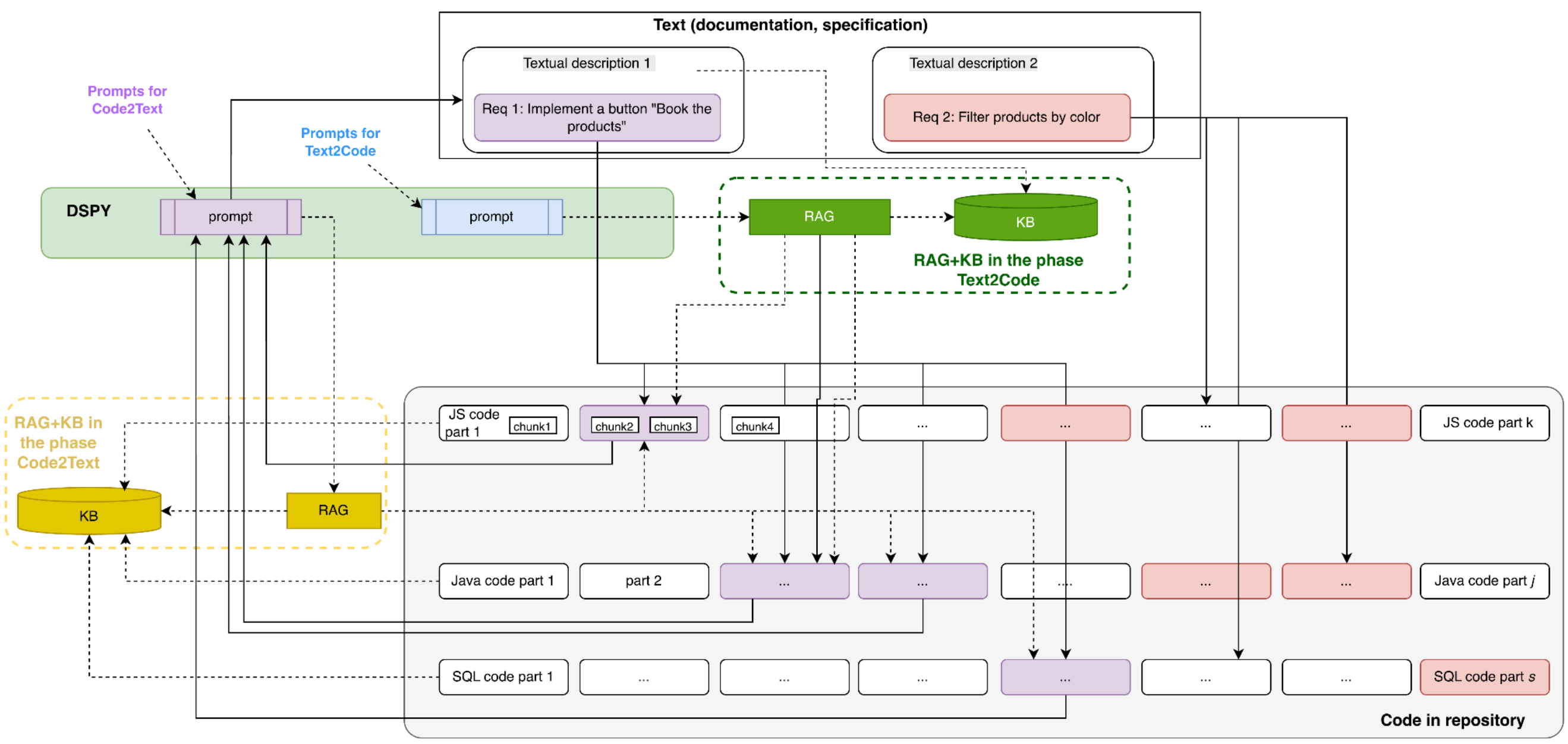

Fig. 7. Text-to-Code mapping with DSPy prompt orchestration, RAG, knowledge base, textual requirements, and cross-language code chunks.

## VII. Experimental Setup

### A. Dataset Construction

The experimental setup combined dataset construction with prompt optimization. DSPy/MIPROv2 was used as a prompt optimization mechanism for testing whether automatically generated instruction variants could improve the baseline prompts [37].

A Code–Text–Code dataset was created to evaluate the proposed pipeline and tune prompts for realistic bidirectional transformation tasks. The source was the GitHub Code dataset, which contains 115 million code files in 32 programming languages [38]. From this corpus, 1000 scripts were selected for each of nine languages: Java, C, C#, Python, C++, JavaScript, PHP, Ruby, and MSSQL. In addition, examples in several SQL dialects were collected from over 30 repositories: BigQuery, PL/SQL, PL/pgSQL, Snowflake SQL, and T-SQL. The selected languages were grouped into stacks: SQL, Web, and General.

The dataset was organized into technology stacks to reflect realistic migration and regeneration scenarios. Table IV summarizes the language groups and their experimental purposes.

TABLE IV. Dataset Composition by Language Stack

| Stack | Languages / dialects | Purpose |
|---|---|---|
| SQL | BigQuery, PL/pgSQL, PL/SQL, Snowflake SQL, SQL, T-SQL | Database migration and dialect transformation |
| Web | PHP, JavaScript, Python | Web-oriented code transformation |
| General | Python, Java, C++, C, C#, Ruby | General programming-language transformation |
| Project documentation subset | SQL and mixed-language projects | Code-to-project-documentation evaluation |
| Documentation transformation subset | Technical and business documents | TechDoc–BusinessDoc–TechDoc evaluation |

This organization allows the framework to be evaluated across SQL-specific, web-oriented, and general-purpose programming transformations, while also supporting documentation-oriented experiments.

### B. Data Validation

After the automatic generation of source-target pairs, the instances were manually checked by domain experts. If errors were found in the source code, the corresponding instances were removed from the dataset. This step was necessary to avoid evaluating the transformation pipeline on invalid or misleading examples.

### C. Prompt Tuning

DSPy and MIPROv2 were investigated for optimizing the prompts [37]. MIPROv2 generates examples, creates instruction variants, runs candidates on test examples, compares them on a selected metric, refines the candidates, and selects the prompt that consistently exhibits the highest quality. As a result, none of the generated candidates outperformed the original prompt. Therefore, the original prompt was retained. The likely reasons were that the base prompt already contained sufficient constraints, including inference structure, neutrality, and identifier preservation; alternative prompts did not provide statistically significant improvement; and paraphrasing made the instruction more difficult without increasing relevance.

### D. Evaluation Design

The evaluation design included comparison of direct text conversion and several intermediate representation strategies: pseudocode, natural language, graph-based IR, and compiler-style IR. Outputs were compared against ground truth using LLM-based and text-similarity metrics. This design supports the empirical selection of neutral textual representation as the central intermediate layer.

To compare intermediate representations, the evaluation pipeline contrasts direct text conversion with pseudocode, natural-language IR, graph-based IR, and compiler-style IR. The workflow is shown in Fig. 8.

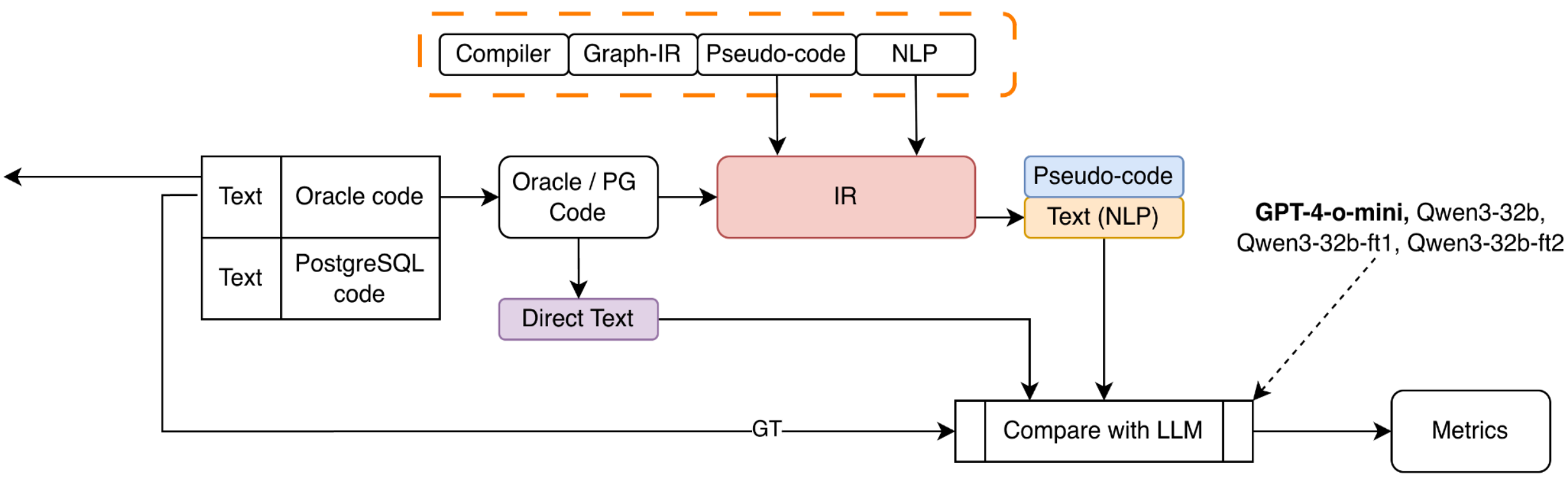

Fig. 8. Intermediate representation evaluation workflow comparing direct text, pseudocode, natural-language IR, graph IR, and compiler IR against ground truth using LLM-based metrics.

This evaluation workflow is important because it separates the effect of the intermediate representation from the effect of the final model response. It therefore supports the empirical selection of neutral textual specification as the central representation layer.

## VIII. Evaluation Results

### A. Intermediate Representation Comparison

The comparison of intermediate representations showed that natural-language IR is highly suitable for LLM-mediated transformation. Fig. 8 presents the evaluation workflow used to compare direct text conversion with pseudocode, natural-language IR, graph-based IR, and compiler-style IR. The outputs were compared against ground truth using LLM-based and text-similarity metrics. The main experimental insight is that the Code-to-Natural-Language pipeline achieved the strongest overall performance among the investigated alternatives. This confirms the role of neutral textual specification as the central intermediate representation. At the same time, the result does not imply that graph or AST representations are unnecessary; rather, they should be used as evidence and traceability layers supporting the textual specification.

In the tested configuration, the fine-tuned Qwen-32B model underperformed because it had been fine-tuned for a different objective; therefore, task-specific fine-tuning must be aligned with the intended Code–Text–Code transformation target.

### B. Retrieval Evaluation

Retrieval experiments compared vector retrieval and graph retrieval. The retrieval experiments showed that vector retrieval provides greater accuracy and generalization, while graph retrieval provides lower latency and relation-based access. The best-performing embedding model was identified as nomic-ai/CodeRankEmbed because it maintained high accuracy in both code-to-description and description-to-code directions, performed well across SQL and Python, and offered efficient inference time suitable for interactive systems. The vector-vs-graph comparison supports hybrid retrieval: vector retrieval should be used when semantic similarity and relevance are critical, while graph retrieval should be used when dependency navigation and fast relation-based access are required.

### C. Code-to-Project Documentation

The Code-to-Project Documentation evaluation demonstrated that the approach can scale from file-level understanding to project-level documentation. The evaluated flow included file extraction, graph extraction, graph decomposition, cluster documentation, and project synthesis. This supports the claim that Code-Text-Code reengineering is not limited to snippet-level transformation but can support project-level representation and documentation synthesis.

### D. Documentation Transformation Evaluation

The TechDoc-to-BusinessDoc evaluation verified whether business documents remained grounded in technical documentation without hallucinations or critical omissions. The reported metrics were Context Recall = 0.99, Faithfulness = 0.99, and Context Entity Recall = 0.52. The lower entity recall should not be interpreted as a direct failure, because business documentation intentionally abstracts, smooths, and aggregates technical entities, mentioning only those that are relevant at the business level. In the chunk-level evaluation, 24 chunks were assessed: 22 passed and 2 failed. The failed cases included one chunk that distorted inventory update logic and another that contradicted the previous data warehouse report scope and omitted an important report. These results show that documentation transformation is feasible, but they also reveal typical semantic risks that require validation loops and coverage-based evaluation.

### *E. Business-to-Technical and Technical-to-Code Evolution*

The BusinessDoc-to-TechDoc pipeline takes the original business document, business changes, and original technical documentation as input and produces a decision on whether a document needs changes, a change description, a reason for change, and updated technical documentation. The TechDoc-to-Code pipeline maps changed technical documentation to source code, passes changes and original files through an LLM to update code, checks whether the code can be compiled or executed, asks the LLM to update it if needed, and evaluates results. Together, these pipelines provide a backward path from stakeholder-level changes to technical documentation and regenerated code.

### *F. Prompt Tuning Evaluation*

The prompt tuning experiment showed that MIPROv2 did not improve the original prompt. This does not mean that prompt optimization is useless. Rather, it suggests that for constrained Code-Text-Code tasks, a carefully designed basic prompt may already contain the most important requirements: neutrality, structure, identifier preservation, and controlled inference. Additional prompt variants may add complexity without improving performance.

The evaluation combined evidence from intermediate representation comparison, retrieval experiments, chunking analysis, documentation transformation, prompt optimization, and graph-based loss estimation. Table V summarizes these dimensions.

TABLE V. EVALUATION DIMENSIONS AND EVIDENCE

| Evaluation dimension | Evidence source | Observed result / role |
|---|---|---|
| IR suitability | IR comparison | NL IR showed strongest practical suitability |
| Retrieval quality | Vector vs Graph evaluation | Vector more accurate; graph faster |
| Chunk quality | Chunking metrics | Self-parsability and boundary quality are critical |
| Documentation transformation | TechDoc-to-BusinessDoc evaluation | 22/24 chunks passed; 2 failed |
| Prompt optimization | DSPy/MIPROv2 | Baseline prompt retained |
| Structural preservation | Graph metrics | Formal loss estimation possible |
| Interface preservation | Graph IO metrics | External behavior can be measured separately |

The results indicate that the framework should be evaluated as a multi-stage transformation process. Local correctness of generated code is not sufficient; semantic coverage, retrieval grounding, documentation consistency, and structural preservation are also required.

## IX. FORMAL METRICS AND TRANSFORMATION-LOSS ESTIMATION

### *A. Graph-Based Representation of Source and Target Systems*

To estimate transformation losses, the source and target systems can be represented as directed graphs. Let the source system be:

$$A = (V_A, E_A), \quad (4)$$

where $V_A$ is the set of nodes and $E_A$ is the set of directed dependency edges. Nodes may represent modules, functions, procedures, database tables, views, entry points, output artifacts, or services. Edges may represent relations such as READS, CALLS, MODIFIED_BY, WRITES, DEPENDS_ON, or INVOKES.

The target system is represented as:

$$B = (V_B, E_B). \quad (5)$$

This formalization is consistent with the provided mathematical model, where the old and new codebases are represented as directed graphs with component nodes and dependency edges. The graph-based view used for estimating transformation loss is summarized in Fig. 9. It compares the source system graph and the target system graph through structural preservation, reverse compatibility, and interface preservation.

This model complements textual validation by adding a structural measure of transformation quality. It is especially useful when regenerated code is behaviorally plausible but introduces unexpected dependencies, removes required components, or changes external interfaces.

### *B. Directional Structural Preservation*

A graph homomorphism $h: A \to B$ maps nodes of the source graph to nodes of the target graph. If every edge $(u, v) \in E_A$ is preserved as $(h(u), h(v)) \in E_B$, the migration preserves source dependencies.

$$\alpha = 1 - \frac{|E_A^{viol}|}{E_A}, \quad (6)$$

where $E_A^{viol}$ is the set of source edges that are not preserved in the target graph. A high α means that the old system structure is well preserved in the new system. A low α indicates structural loss or broken dependency preservation.

### *C. Reverse Compatibility and Growth Control*

The reverse direction measures how much the target system avoids introducing unmatched structure:

$$\beta = 1 - \frac{|E_B^{viol}|}{E_B}, \quad (7)$$

where $E_B^{viol}$ is the set of target edges that cannot be mapped back to the source graph. A high β means that the target system does not introduce many unnecessary dependencies. A low β may indicate excessive restructuring, unexpected generated behavior, or architectural drift.

### *D. Bidirectional Similarity*

The source-to-target and target-to-source scores can be aggregated using a harmonic mean:

$$H(\alpha, \beta) = \frac{2\alpha\beta}{\alpha+\beta}. \quad (8)$$

A weighted harmonic mean can also be used:

$$H_\gamma(\alpha, \beta) = \frac{1}{\frac{\gamma}{\alpha} + \frac{1-\gamma}{\beta}}, \quad (9)$$

where $\gamma \in [0, 1]$ controls the relative importance of source-to-target preservation and target-to-source compatibility.

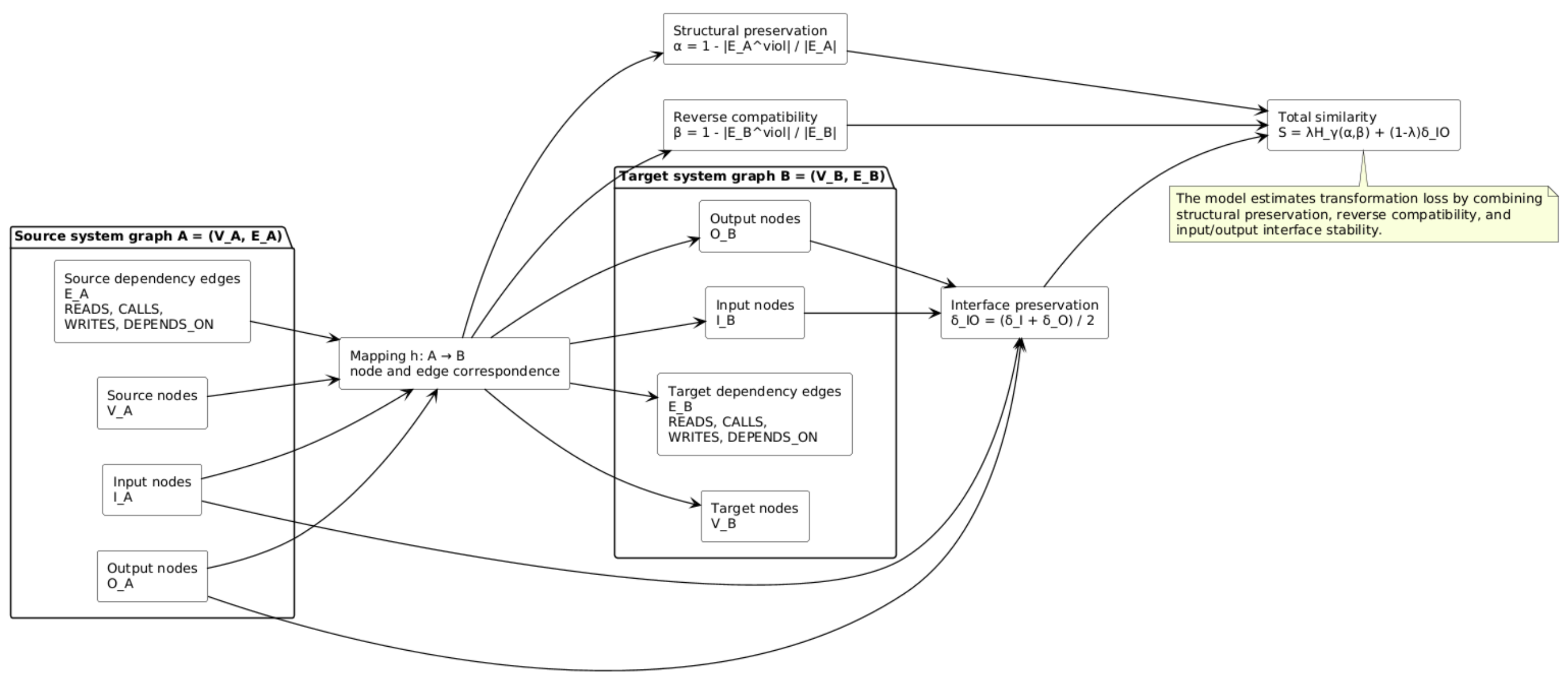

Fig. 9. Graph-based transformation-loss estimation model using source graph A, target graph B, mapping h, structural preservation α, reverse compatibility β, interface preservation δIO, and total similarity S.

The weighted version may emphasize one direction depending on the goal of the transformation. For migration, source preservation may be more important. For modernization, controlled restructuring may be acceptable.

### *E. Interface Preservation*

Let $I_A, O_A \subseteq V_A$ be input and output nodes of the source system, and $I_B, O_B \subseteq V_B$ be input and output nodes of the target system.

$$\delta_I = 1 - \frac{|h(I_A)\Delta I_B|}{I_A \cup I_B},$$

$$\delta_O = 1 - \frac{|h(O_A)\Delta O_B|}{O_A \cup O_B},$$

$$\delta_{IO} = \frac{\delta_I + \delta_O}{2}. \quad (10)$$

A high $\delta_{IO}$ means that external input and output interfaces are preserved.

### *F. Total Graph Similarity*

Total graph similarity can be computed as:

$$S = \lambda H_\gamma(\alpha, \beta) + (1 - \lambda)\delta_{IO}, \quad (11)$$

where $\lambda \in [0, 1]$ balances structural similarity and interface preservation. A higher λ prioritizes internal dependency preservation, while a lower λ prioritizes external compatibility. This follows the formalization where total graph similarity is defined as a weighted combination of structural similarity and interface similarity.

### *G. Additional Evaluation Metrics*

Code–Text–Code reengineering should be evaluated using a combination of syntactic, semantic, retrieval, documentation, and graph metrics. To operationalize the evaluation protocol, the framework combines syntactic, functional, semantic, retrieval, chunking, and graph-based metrics. Table VI presents the corresponding metric set.

The proposed metric set supports both local and system-level evaluation. Syntactic validity and functional equivalence evaluate generated artifacts, while semantic coverage, hallucination rate, retrieval groundedness, and graph similarity evaluate transformation quality across representation layers.

TABLE VI. Formal Metrics for Code–Text–Code Reengineering

| Metric | Formula / definition | Interpretation |
|---|---|---|
| Syntactic Validity Rate | valid outputs / all outputs | Generated artifact is parseable or executable |
| Functional Equivalence | passed behavioral tests / all tests | Target code preserves expected behavior |
| Semantic Coverage | covered source facts / all source facts | Specification covers source semantics |
| Hallucination Rate | unsupported statements / all statements | Measures unsupported generated content |
| Retrieval Groundedness | supported claims / all claims | Measures evidence support |
| Chunk Certainty | parsable chunks / all chunks | Measures chunk syntactic quality |
| Boundary F1 | boundary precision/recall F1 | Measures chunk segmentation quality |
| α | $1 - \|E_A^{viol}\|/\|E_A\|$ | Source-structure preservation |
| β | $1 - \|E_B^{viol}\|/\|E_B\|$ | Reverse compatibility / growth control |
| $\delta_{IO}$ | $(\delta_I + \delta_O)/2$ | Input/output stability |
| S | $\lambda H_\gamma + (1 - \lambda)\delta_{IO}$ | Total transformation similarity |

## X. Results and Discussion

To connect the proposed framework with the evidence presented in the paper, Table VII maps each research contribution to the corresponding sections, figures, and evaluation results. This mapping clarifies that the paper contributes not only an architectural pipeline, but also a knowledge representation model, an evaluation protocol, and a formal loss-estimation mechanism for LLM-mediated software evolution.

### *A. Why Code–Text–Code Instead of Direct Code–Code?*

Direct code-to-code conversion is attractive for its simplicity. However, it hides the interpretation of the model. The user sees only the source and the generated target, but not the intermediate considerations or semantic representation. This makes it difficult to determine whether the model has preserved the behavior or simply generated plausible code.

TABLE VII. RESEARCH CONTRIBUTIONS VS EVIDENCE

| Contribution | Description | Evidence in this paper |
|---|---|---|
| C1. Conceptual model | Code–Text–Code interpreted as specification-based reengineering | Sections I and III; Fig. 1 |
| C2. Ontology-supported knowledge model | AST, metadata, ontology, graph, neutral text, documentation, target code | Section IV; Fig. 2; Table II |
| C3. Pipeline architecture | Code processing, IR, ERM, RAG, GraphRAG, Text2Code | Section V; Fig. 3 |
| C4. Container-level implementation | UI, backend, metadata extraction, vector DB, RAG, LLM gateway, validation gateway | Section V; Fig. 4 |
| C5. DBMS-supported conversion | Dialect-aware parsing, DBMS feedback, static/LLM conversion, validation | Section V.F; Fig. 5 |
| C6. Chunking and retrieval | Semantic-aware chunking, RAG-supported Code2Doc, hybrid retrieval | Section VI; Fig. 6; Table III |
| C7. Text-to-Code mapping | DSPy, RAG, KB, textual requirements, cross-language chunks | Section VI.E; Fig. 7 |
| C8. Experimental grounding | Dataset, IR comparison, retrieval, TechDoc–BusinessDoc evaluation | Sections VII–VIII; Fig. 8; Tables IV–V |
| C9. Formal loss metrics | Graph-based α, β, δIO, S | Section IX; Fig. 9; Table VI |

Code-Text-Code introduces an explicit intermediate artifact. The neutral textual specification makes the interpretation of the model visible. It can be reviewed, tested, corrected, and reused. This improves transparency and reduces uncontrolled semantic drift.

### B. *Role of Neutrality*

Neutrality is important. If the intermediate text contains terms specific to the source language, the target code may imitate the constructs of the source. If the text is too abstract, the target code may omit implementation details. If the text contains unsupported assumptions, the target code may implement behavior that was never present in the source code. Therefore, the specification should be sufficiently detailed, but not syntactically biased.

### C. *Traceability as a Control Mechanism*

Traceability links source code, prompts, metadata, graph nodes, specification text statements, documentation sections, and generated code. Without tracing, validation becomes subjective. With tracing, each generated artifact can be checked for evidence.

### D. *Retrieval and External Knowledge*

Many code chunks use libraries, frameworks, APIs, database functions, or platform-specific constructs. The model may not correctly interpret them without external context. RAG can provide documentation and examples, but retrieval must be grounded. Metadata and graph representation can help retrieve more precise context than raw text alone.

### E. *Human-in-the-Loop Role*

This framework includes optional human involvement, as not all ambiguities can be resolved automatically. Human experts can clarify business meaning, validate assumptions, resolve name ambiguities, or approve behavioral changes. This is especially important in high-risk systems, migration projects, financial systems, healthcare software, and corporate reporting.

### F. *Multi-Agent Extension*

A further extension may involve multi-agent orchestration, where different agents specialize in source-code analysis, metadata extraction, graph construction, ontology reasoning, specification generation, retrieval, validation, transformation-loss estimation, and target-code regeneration. Recent work on Semantic Kernel orchestration shows that dynamic involvement of specialized agents can support complex tasks requiring adaptive coordination [39]. This extension is not required for the basic Code–Text–Code pipeline, but it becomes important when the framework is scaled from isolated transformations to project-level workflows involving separate agents for parsing, retrieval, validation, ontology reasoning, and code regeneration.

## XI. VALIDATION PROTOCOL

### A. *Internal Validity*

The evaluation relies in part on LLM-as-judge judgments and transformation results. Such judgments may be inconsistent across model versions, queries, and temperature settings. To mitigate this threat, validation should combine LLM-as-judge results with deterministic checks, parser validation, graph comparison, test execution, and expert review.

### B. *Construct Validity*

Metrics such as semantic coverage, retrieval groundedness, hallucination coefficient, and functional equivalence only approximate the true correctness of code-text-code transformations. A target artifact may pass syntactic validation but still fail to preserve implicit assumptions. Conversely, a structurally different target implementation may still be behaviorally correct. Therefore, the interpretation of metrics must be contextual.

### C. *External Validity*

The dataset includes multiple programming languages and SQL dialects, but the results may not generalize to all domains, especially safety-critical systems, embedded software, real-time systems, legacy mainframe systems, or highly dynamic distributed architectures. Additional experiments with industrial codebases and domain-specific repositories are needed.

### D. *Conclusion Validity*

Some published experimental results are based on limited validation samples. For example, the MIPROv2 prompt tuning did not outperform the baseline prompt, but this may depend on the sample size, problem formulation, chosen metric, and model family. Larger experiments are needed to confirm the generality of this observation.

### E. *Tooling Validity*

AST extraction, graph construction, chunking, and metadata quality depend on parsers, dialect support, and static analysis tools. SQL dialects are particularly complex because parser support varies depending on dialect specifics. Tool failures can extend to metadata, graph representation, retrieval, and validation.

## CONCLUSIONS

This paper proposed a framework for specification-based Code-Text-Code reengineering for LLM-mediated software evolution. The main argument is that LLM-assisted software transformation should not be interpreted as direct code generation or code translation alone. Instead, it should be treated as a controlled reengineering process in which source code is transformed into a neutral textual specification and then regenerated into target code.

The proposed framework integrates factual hint extraction, Code2Text generation, iterative validation, human-in-the-loop correction, Text2Code generation, target-code validation, metadata extraction, ontology-based knowledge representation, graph-based traceability, hybrid retrieval, documentation transformation, semantic-aware chunking, dialect-aware parsing, DBMS feedback, and graph-based transformation-loss estimation. The neutral textual specification acts as the central knowledge representation layer that links source code, technical documentation, business documentation, and target code.

Experimental evidence supports the feasibility of the approach. Textual intermediate representation demonstrated strong practical suitability as an intermediate layer. Metadata and graph representations support traceability. Hybrid retrieval balances semantic relevance and structural navigation. Documentation transformation experiments show promising results but also reveal semantic risks such as logic distortion and scope omission. Prompt tuning results indicate that carefully constrained baseline prompts can outperform more complex optimized variants in some Code-Text-Code scenarios.

The proposed graph-based formalization provides a way to estimate transformation losses by measuring structural preservation, reverse compatibility, interface stability, and total graph similarity. This makes the framework relevant not only for code generation, but also for intelligent monitoring of software evolution. Overall, specification-based Code-Text-Code reengineering provides a promising direction for moving from opaque direct generation toward traceable, verifiable, and monitorable transformation across representation layers.